\documentclass[aps,prb,twocolumn,superscriptaddress,showpacs]{revtex4-1}
\usepackage{graphicx}
\usepackage[german,english]{babel}
\usepackage{amsmath}

\begin{document}

\title{Adhesion and electronic structure of graphene on hexagonal boron nitride substrates}
\author{B. Sachs}
\email{bsachs@physnet.uni-hamburg.de}
\affiliation{I. Institut f{\"u}r Theoretische Physik, Universit{\"a}t Hamburg, Jungiusstra{\ss}e 9, D-20355 Hamburg, Germany}

\author{T. O. Wehling}
\email{twehling@physnet.uni-hamburg.de}
\affiliation{I. Institut f{\"u}r Theoretische Physik, Universit{\"a}t Hamburg, Jungiusstra{\ss}e 9, D-20355 Hamburg, Germany}

\author{M. I. Katsnelson}
\affiliation{Radboud University of Nijmegen, Institute for
Molecules and Materials, Heijendaalseweg 135, 6525 AJ Nijmegen,
The Netherlands}

\author{A. I. Lichtenstein}
\affiliation{I. Institut f{\"u}r Theoretische Physik,
Universit{\"a}t Hamburg, Jungiusstra{\ss}e 9, D-20355 Hamburg,
Germany}

\pacs{81.05.ue; 73.22.Pr; 71.10.--w; 71.20.--b}


\date{\today}

\begin{abstract}
We investigate the adsorption of graphene sheets on
h-BN substrates by means of first-principles calculations in the framework of adiabatic connection fluctuation-dissipation theory in the random phase approximation. We obtain adhesion energies for different crystallographic stacking configurations and
show that the interlayer bonding is due to long-range van der Waals forces. The interplay of elastic and adhesion energies is shown to lead to stacking disorder and moir\'e structures. Band structure calculations reveal substrate induced mass terms in graphene which change their sign with the stacking configuration. The dispersion, absolute band gaps and the real space shape of the low energy electronic states in the moir\'e structures are discussed. We find that the absolute band gaps in the moir\'e structures are at least an order of magnitude smaller than the maximum local values of the mass term. Our results are in agreement with recent STM experiments.
\end{abstract}
\maketitle

\section{Introduction}

The development of graphene-based nanoelectronic devices, such as high-speed transistors, calls for high electron mobilities. Currently, substrates beneath the graphene present an important source of disorder including corrugation effects of the graphene \cite{Fuhrer07,Morgenstern09,ripples1,ripples2}, charge traps \cite{Fuhrer07,chargedimp1,chargedimp2,chargedimp3,chargedimp4,chargedimp5} and dangling bonds \cite{dangling1,dangling2}. A promising candidate to become a new standard substrate material is hexagonal boron nitride (h-BN). This BN polymorph is remarkably similar to graphite: The alternating B and N atoms form two-dimensional layers of strong sp$^2$ bonds within a honeycomb arrangement and a lattice constant which differs by less than 2\% from that of graphene. The h-BN sheets are weakly bound by long-range adhesive forces at an equilibrium distance of 3.3\AA~\cite{hBNlayersep}. The electronic structure, however, exhibits clear differences: the chemically inequivalent sublattices make h-BN an insulator with a band gap of 6.0 eV \cite{hBNgap}. Recently, the fabrication of graphene devices on h-BN with highly improved electron mobilities and carrier inhomogeneities, reaching a quality comparable to suspended graphene has been reported \cite{hBNmain,hBNtrans}. Thereby, the graphene was found to keep its zero band gap and to stack quasi randomly orientated on the h-BN substrate.

In this paper, we analyze the adhesion behavior and the electronic structure of graphene on h-BN from first-principles. The paper is organized as follows: In section \ref{sec:adhesion}, we show that methods beyond standard density functional theory (DFT) are necessary to describe the weak non-local attraction between the h-BN and the graphene layers. We calculate adhesion energies using the random phase approximation (RPA) within the framework of the adiabatic connection fluctuation-dissipation theorem (ACFDT). On the basis of elasticity calculations, we discuss in section \ref{sec:moire} mechanisms to release the stress resulting from the lattice mismatch and leading to the formation of moir{\'e} superstructures. Section \ref{sec:band_str} is devoted to the band structure and energy gaps of graphene on h-BN. From DFT band structure calculations we derive a low energy tight-binding description of graphene on h-BN and find mass terms which change their sign with the stacking configuration. This leads to an absolute gap in the moir\'e structure which is at least an order of magnitude smaller than the maximum local values of the mass term. The real space shape of the low energy states, particularly the issue of sublattice polarization and the occurrence of so-called snake states in regions where mass term changes its sign is discussed in section \ref{sec:real_space}. Finally, conclusions and an outlook are given in section \ref{sec:concl}.

\section{Adhesion of graphene on \lowercase{h}-BN}
\label{sec:adhesion}

DFT is a successful approach to describe ground state properties of solids. However, widely used semilocal approximations for the exchange-correlation energy like the local density approximation (LDA) and generalized gradient approximation (GGA) \cite{LDAGGA1,LDAGGA2} do not take long-range correlations into account correctly. Thus, they fail to reproduce van der Waals attraction and the prediction of equilibrium geometries of van der Waals bound layered systems, such as graphite or h-BN, proves problematic with these methods \cite{Rubio06,hBNgraphitef2}. A highly accurate means to describe van der Waals forces from first-principles is provided by the random phase approximation to the correlation energy \cite{RPA1,RPA2}. Evaluated in the ACFDT framework, the RPA correlation energy reads \cite{RPA_theo}
\begin{equation}
E_c^{\rm RPA}=\int_0^{\infty} \frac{d\omega}{2\pi} \mathrm{Tr}\{ \ln[1-\nu \chi^{KS}(i\omega)] + \nu \chi^{KS}(i\omega)\},
\end{equation}
where $\chi^{\rm KS}$ is the response function of the non-interacting Kohn-Sham (KS) system and $\nu=\sum_{i < j}\frac{e^2}{\mid \vec{r}_i -\vec{r}_j \mid}$ the electron-electron interaction. Together with the total KS Hartree-Fock energy, usually referred to as exact exchange energy $E_{\rm EXX}$ \cite{EC_RPA}, the total ground-state energy of the system reads as $E = E_{\rm EXX} + E_c^{\rm RPA}$. The ACFDT-RPA method has been proven accurate to describe bulk properties of solids such as lattice constants as well as adsorption energies \cite{EC_RPA}. For layered van der Waals bonded systems ACFDT-RPA yields a much more accurate description of the structural properties than LDA and vdW DFT methods \cite{graphitekresse,hBNrubio}. 

\begin{figure}[bt]
	\centering
	\includegraphics[width=1.0\linewidth]{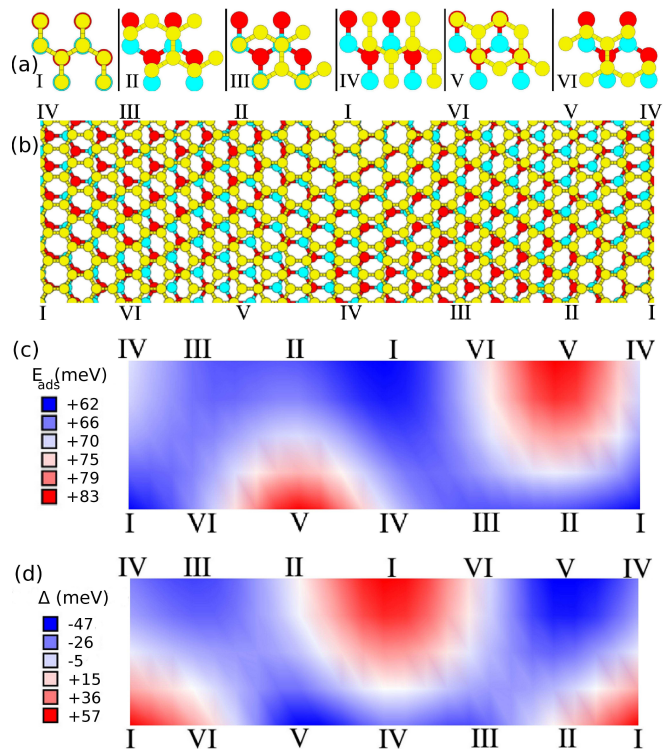} 
\caption{(Color online) (a) Top view of the calculated stacking configurations for graphene on h-BN with the carbon atoms (yellow), boron (red) and nitrogen (light blue). Between each neighboring images the graphene lattice is shifted downwards by half a B-N bond. (b) Moir\'e structure with persisting lattice mismatch. A lattice mismatch of 1.8\% corresponds to a $55\times 55$ moir\'e unit cell. For clarity, a smaller moir\'e unit cell (13x13) is shown. (c) Adhesion energy ($E_{\rm ads}$) landscape in the moir\'e pattern (color-coded). (d) The same for the local sublattice symmetry breaking $\Delta$.} 
\label{fig:confs} 
\end{figure}

To simulate the graphene--h-BN system in this way, calculations were performed with the Vienna \textit{ab initio} simulation package (VASP) \cite{VASP} using plane wave basis sets within the projector-augmented wave (PAW) method \cite{PAW1,PAW2}. To this end, a unit cell (4 atoms) containing graphene on an h-BN layer with 25\AA~ of vacuum above was constructed. Six stacking  configurations were considered (Fig. \ref{fig:confs}a): starting from configuration I, the graphene sheet was translated downwards by half a B-N bond length in each step until the initial configuration was reached again. The lattice constant was chosen to $2.49$\AA, referring to the LDA optimized lattice constant of h-BN (a discussion of the lattice mismatch follows below). $\chi^{KS}$, $E_c^{\rm RPA}$ and $E_{\rm EXX}$ were evaluated with the LDA KS orbitals. In these computationally demanding simulations, convergence of the results was reached at a kinetic energy cut-off of 347 eV for the response function, a plane wave cut-off of 520 eV and a mesh of $7\times 7\times 1$ k-points. Additionally, standard LDA/GGA calculations were performed for the same geometries with a $24\times 24\times 1$ k-points grid and a kinetic energy cut-off of 500 eV.

\begin{figure}
	\centering
	\includegraphics[width=1.0\linewidth]{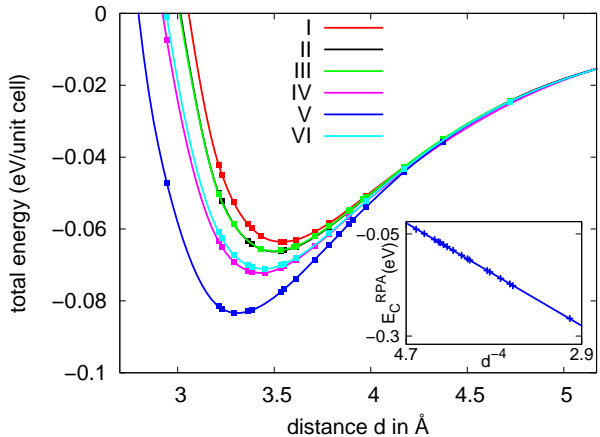} 
\caption{(Color online) Total RPA energies per unit cell for the stacking configurations I--VI as a function of the distance, $d$, between the graphene and h-BN layer. Inset: RPA correlation energy (relative to the correlation energy at large separation) of configuration V as function of $d^{-4}$ between 2.9\AA~ and 4.7\AA.}
\label{fig:RPA6conf}
\end{figure}

Fig. \ref{fig:RPA6conf} shows the RPA total energies per unit cell (relative to the energy at large separation) for the different stacking configurations as a function of the interlayer spacing $d$. The curves show that, starting from the highest-energy configuration I, the lowest-energy configuration V is approached step-wise. 
The configurations I, II and III exhibit total energy minima at interlayer spacings between 3.50\AA~ and 3.55\AA, whereat II and III are energetically virtually equivalent with minima of -65 meV; the minimum  of configuration I is only slightly higher ($-62$\,meV). Configuration V, where the carbon atoms sit on top of a boron atom and in the middle of the BN hexagon, is energetically most favorable with a minimum of -83 meV and an equilibrium layer distance of 3.35\AA. Energetically closest to V are IV (-71 meV) and VI (-70 meV), where the nitrogen atom is also not covered by a C atom or a C-C bond. The curves exhibit no additional energy barrier for translation. For distances larger than 4\AA, all configurations become energetically indistinguishable. 

To get further insight to the nature of the attractive forces between graphene and h-BN, we analyze the decay of the RPA correlation energy $E_{c}^{\rm RPA}$ with the interlayer spacing $d$ (Fig. \ref{fig:RPA6conf} inset). We find $E_{c}^{\rm RPA} \sim d^{-4}$. This is clearly different from an exponential falloff as would be expected for local correlation effects as included in LDA or GGA. It rather indicates bonding of vdW type which yields power law decays\cite{Rubio06} $E_{c}^{\rm RPA} \sim d^{-p}$ with $p=4$ for 2D insulating systems \cite{hBNgraphitef2,Rubio06}.


\begin{figure}
	\centering
	\includegraphics[width=1.0\linewidth]{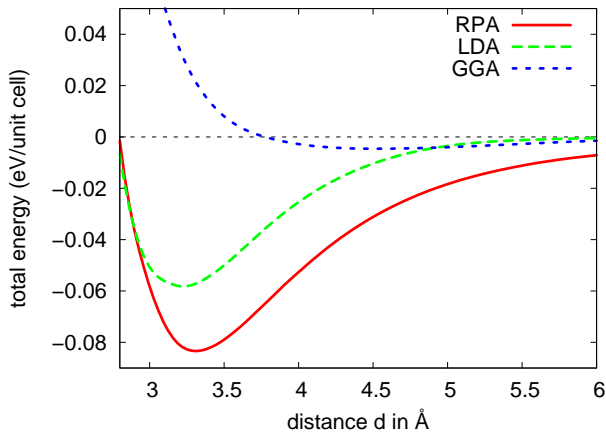}
\caption{(Color online) Total energies per unit cell of configuration V from RPA, LDA and GGA.}
\label{fig:RPAGGALDA}
\end{figure}

A comparison of the RPA calculations with the standard LDA and GGA methods is given in Fig. \ref{fig:RPAGGALDA}. The LDA (see also \onlinecite{khomyakov}) yields a qualitatively correct equilibrium layer separation but underestimates the RPA adhesion energy by about 30\%.  The GGA, actually an improvement over the LDA in many cases \cite{LDAGGA1,LDAGGA2}, exhibits an even more dramatic underbinding with a weak minimum giving a negligible adhesion energy of less than 6 meV/unit cell at an extremely high equilibrium distance. The LDA and GGA curves nearly coincide for distances larger than 4.5\AA~and underestimate the van der Waals interaction also in this asymptotic region. The results clearly show that long-range correlations have to be taken into account for an accurate description of the layer attraction.

\section{Existence of moir\'{e} structures}
\label{sec:moire}

We now turn to the discussion of the consequences of the lattice mismatch (1.8\% in LDA and 1.9\% in GGA) between the graphene and the h-BN and address the question whether stacking disorder or moir{\'e} superstructures should occur. To this end, we estimate the total energy difference of a structure with stacking according to the minimum energy configuration V in the entire sample and a moir\'e structure with persisting mismatch (Fig. \ref{fig:confs}b). 

In the case of persisting mismatch in the system, a moir\'e pattern with a large unit cell (55x55 for 1.8\% mismatch) is formed (Fig. \ref{fig:confs}b). Here, out-of-plane corrugations resulting from interlayer spacings varying between 3.35\AA~ (region V) and 3.55\AA~ (region I) are negligible, since their amplitude (0.2\AA) is small as compared to their wavelength ($\sim$135\AA). Hence, we focus on in-plane deformations. It is visible from Fig. \ref{fig:confs}b that our choice of stacking configurations I--VI simulated in RPA covers the sample uniformly and gives a sketch of the energy landscape (Fig. \ref{fig:confs}c). Those parts of the moir\'e pattern, where the nitrogen atoms are mainly beneath the center of the carbon rings (regions IV--VI), are energetically more favorable than regions I--III. 
For the moir{\'e} structure we estimate the average adhesion energy per two carbon atoms by the average over the adhesion energies of the configurations I-VI. We obtain an average adhesion energy of 69 meV/(2 C-atoms). This is 14 meV/(2 C-atoms) less than the adhesion energy of 83 meV/(2 C-atoms) in configuration V. 

In the other case with the entire sample in configuration V, the lattice mismatch must be overcome and strain energy has to be brought up to force graphene and h-BN to have the same lattice constant. 
Now, two situations have to be distinguished. For graphene on a h-BN crystal as in Refs. \onlinecite{hBNmain, hBN_STM, ponomarenko2011tunable}, the uppermost h-BN layer will likely keep its lattice constant at the bulk value. To stretch the graphene on the lattice constant of h-BN, 40 meV/(2 C-atoms) of strain energy have to be overcome. Therefore, the strain energy overcompensates the adhesion energy gain of 14 meV/(2 C-atoms) by far when forcing the entire sample to configuration V. Hence, the lattice mismatch between the graphene and the h-BN will persist and strain will be released by realizing different stacking configurations as in the moir{\'e} structure depicted in Fig. \ref{fig:confs}. This explains why multiple stacking configurations have been realized in the experiment of Ref. \onlinecite{hBNmain} and also explains the recent observations of moir\'e patterns in STM experiments \cite{hBN_STM}. 

Second, one can conceive a situation where graphene is adsorbed on a free-standing monolayer of h-BN. In this case, the elastic properties of h-BN have to be accounted for since, then, a compression of h-BN is possible. Our first-principles calculations (see App. \ref{sec:elasticity}) yield two-dimensional Lam\'e parameters and Young's moduli of $\lambda=59$N/m, $\mu=125$N/m, $Y_{\rm h-BN,2D}=309$N/m for single h-BN sheets. Compared to graphene, where $Y_{\rm G,2D} \approx 340$N/m ($Y_{\rm G,3D} \approx 1.0$TPa) \cite{grapheneyoung}, the stiffness of h-BN is on the same order (about 10\% smaller) and thus remarkably high. For graphene on free-standing h-BN, we find that a composition of stretched graphene and compressed h-BN is energetically most favorable and obtain a common optimized lattice constant of 2.467\AA~(LDA) with the total strain energy being 18 meV/(2 C-atoms). Hence, this strain energy is very close to the adhesion energy gain of 14 meV/(2 C-atoms). This might lead to an interesting competition of these two energy contributions and one might expect that systems with graphene on free-standing h-BN are highly sensitive to the experimental environment.

\section{Band structure and energy gaps}
\label{sec:band_str}

We now investigate the band structure of graphene--h-BN hybrid structures and study the changes upon formation of a moir\'e structure. To this end, we calculated the band structure of all geometries depicted in Fig. \ref{fig:confs}a within the LDA. For configurations II, IV and VI we detect a small shift of the Dirac point in the hexagonal Brillouin zone away from the K to the M point (IV, VI) and in the opposite direction (II). Mapping the problem on a nearest-neighbor tight-binding model, we see that a description of graphene in configurations II, IV and VI requires two different hopping parameters, $t$ and $\tilde{t}$, as the threefold symmetry of the graphene nearest neighbor bonds is broken --- analogous to the case of uniaxially strained graphene \cite{ripples1,ripples2}. Fits of the TB model to the DFT results yield constant $t=2.45$\,eV in all regions and $\tilde{t} \neq t$ in regions II, IV and VI (Table \ref{tab:gap}). In agreement with \cite{khomyakov,zasada_hBN_band}, we extract finite band gaps $\Delta$ in all regions varying between 7 meV and 57 meV. However, we find that these gaps have different signs (Table \ref{tab:gap}), where we use the convention that a $\Delta>0$  corresponds to states close to valence band maximum being entirely localized in sublattice B, while $\Delta <0$ corresponds to states at valence band maximum being localized in sublattice A.

In a moir\'e structure like in Fig. \ref{fig:confs}d this leads to a landscape of local sublattice symmetry breaking $\Delta$ with changing signs (Fig. \ref{fig:confs}d, Table \ref{tab:gap}). We note that the local sublattice symmetry breaking does not necessarily lead to local spectral gaps in the LDOS.
\begin{table}%
\begin{tabular}{|c|c|c|c|c|c|c|}
\hline
& I& II& III& IV& V& VI\\
\hline
$\Delta$(meV) &  +57&  +7&  -34&  -25& -47& +14\\
$1-\tilde{t}/t$ & 0& 0.010& 0& -0.002& 0& -0.010\\
\hline
\end{tabular}
\caption{Band gap $\Delta$ and ratio of the two inequivalent nearest-neighbor hopping parameters $\tilde{t}/t$ with $t=2.45$eV in structures with broken trigonal symmetry. The average band gap is -4 meV.} \label{tab:gap}
\end{table}
To gain understanding of the effect of the modulated gap landscape in the moir\'e structure on graphene electrons, we consider the following tight binding model:

\begin{equation}
\label{eqn:TB}
H= - t \sum_{\langle i,j \rangle} \left( a_{i}^{\dagger} b_{j} + h.c. \right) + \frac{1}{2} \sum_i \Delta_i \left( a_{i}^{\dagger} a_{i} - b_{i}^{\dagger} b_{i} \right)
\end{equation}
with $t=2.45$ eV the nearest-neighbor hopping, $a_{i}^{\dagger}$ ($b_{i}^{\dagger}$) the creation operators of an electron on sublattice A (B), and $\Delta_i$ the local mass term. The lattice vectors of the moir\'e unit cell of size $n \times n$ are defined as $\vec{a}_{n_{1,2}}=n \vec{a}_{1,2}$  with $\vec{a}_{1}= a \left(1,0 \right)$ and $\vec{a}_{2}= a \left( -1/2, \sqrt{3}/2 \right)$ being the simple graphene unit cell vectors and $a$ the graphene lattice constant. We denote local mass term by $\Delta_i$, where $i=(l,m)$ describes the position within the the moir\'e cell. $\Delta_i$
 is periodic with the moir\'e cell.
Transforming the $\Delta_i$ to the reciprocal space, we find that the zeroth order Fourier component 
\begin{equation}
\label{eqn:0thgap}
\Delta_{\vec{G}=0}=\frac{1}{N} \sum_i \Delta_i
\end{equation}
is given by the average of all local gaps in the moir\'e cell. In addition to $\Delta_{\vec{G}=0}$, the effect of the first order components of $\Delta_{\vec{G}}$ with the smallest possible $\vec{G} \neq 0$ on the band structure is discussed in the following. To this end, we consider a sinusoidally modulated gap term 
\begin{equation}
\label{eqn:masses}
\begin{aligned}
\Delta_i = A \sin\left(2 \pi l/n  + \Phi_1 \right) + 
B  \sin\left(2 \pi m/n + \Phi_2 \right) + C.
\end{aligned}
\end{equation}
Here, $A$, $B$, $C$ and $\Phi_{1,2}$ denote constants. Taking the local mass terms obtained from DFT in regions I-VI, \emph{realistic} parameters $A=18.6$ meV, $B=42.0$ meV, $\Phi_1=1.884$ and $\Phi_2=1.531$ can be obtained from a fit. 

\begin{figure*}
\includegraphics[width=0.97\columnwidth]{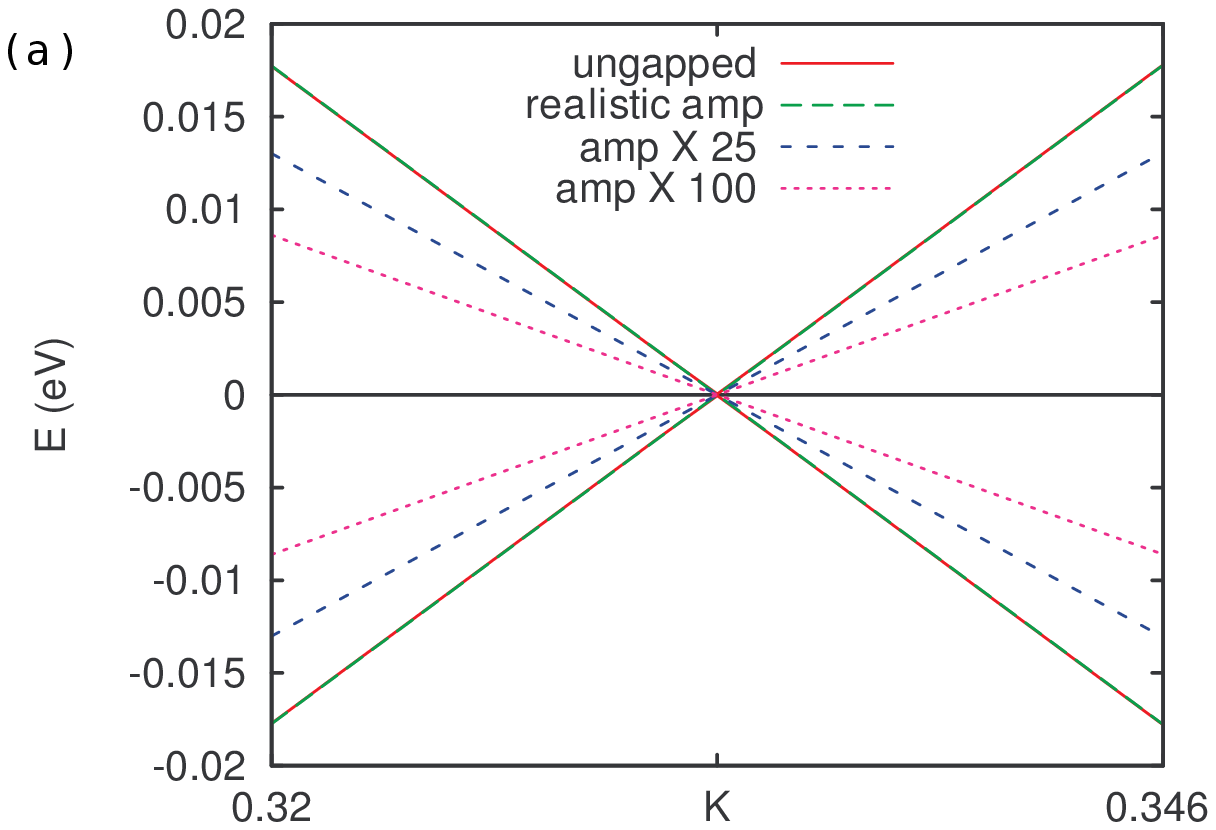}
\includegraphics[width=0.97\columnwidth]{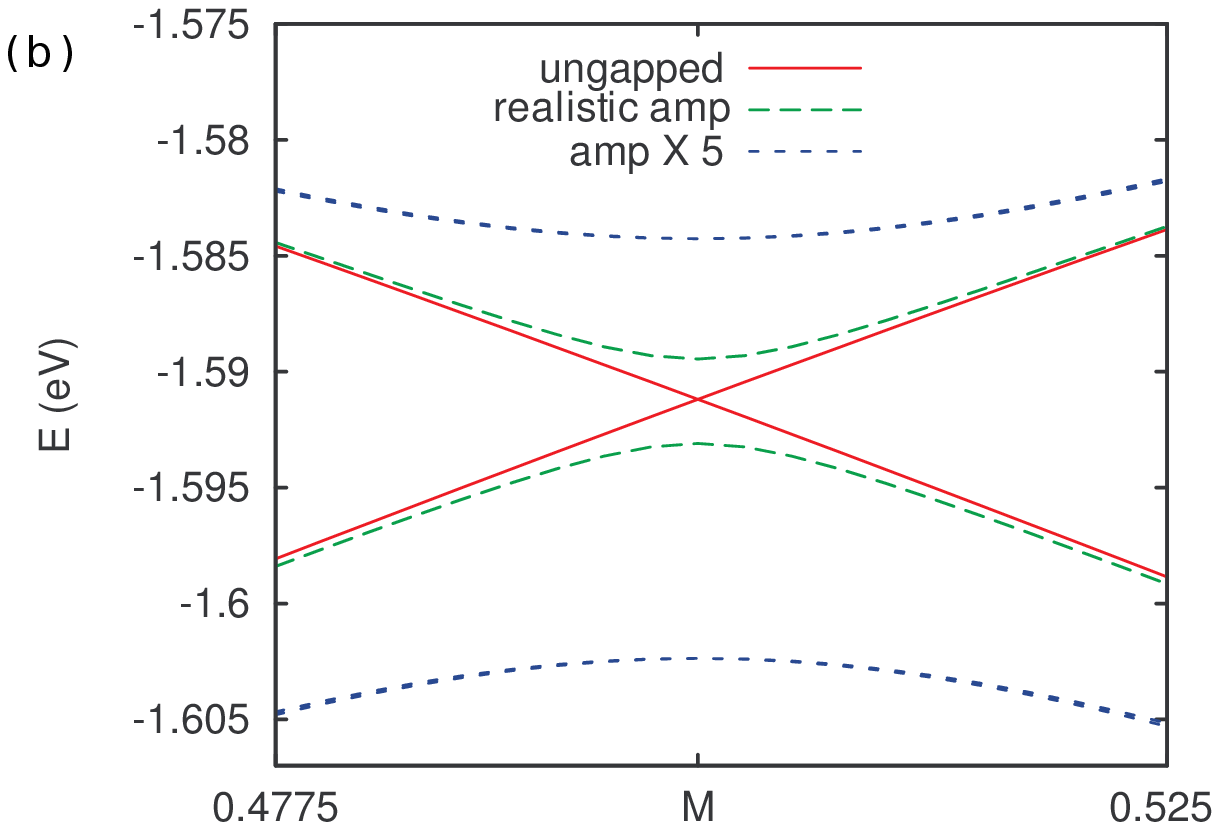}
\caption{\label{fig:bands} {(color online) (a) Bands close to the Fermi level of a 20x20 moir\'e cell of entirely ungapped graphene (red line) and graphene with sinusoidally modulated gap (average gap $\Delta_{\vec{G}=0}=0$) with realistic amplitudes (green dashed line) and increased amplitudes $A$, $B$ (blue/purple dashed lines). No band gap opens. For visualization purposes, the Dirac point has been shifted back to the K point. (b) Bands below the Fermi level at the Brillouin zone boundary (folded to the M point) for ungapped graphene (red line), realistic amplitudes (green dashed line) and increased amplitudes (blue dashed line): minigaps open for finite amplitudes.}} %
\end{figure*}

\begin{figure}
\includegraphics[width=0.97\columnwidth]{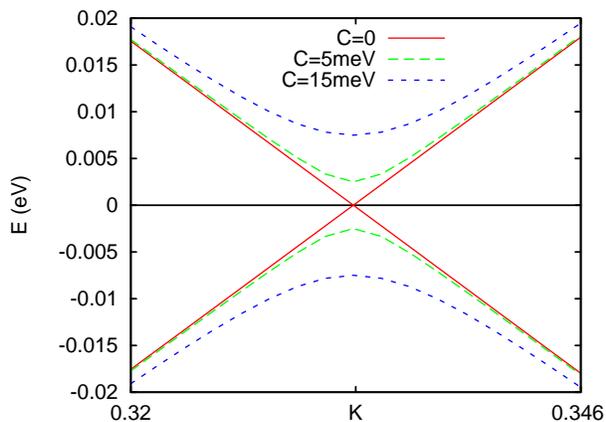}
\caption{\label{fig:bands2} {(color online) Bands close to the Fermi level of a 20x20 moir\'e cell with a sinusoidally modulated gap with realistic amplitudes and non-zero average gap $\Delta_{\vec{G}=0}=C$. Band gaps as large as $C$ open.  }}
\end{figure}

First, we concentrate on the question whether the modulated gap landscape opens an absolute band gap or not. In Fig. \ref{fig:bands}a, the two energy bands closest to the Fermi level for a 20x20 graphene supercell are depicted with a gap landscape as given by Eq. \ref{eqn:masses} and different amplitudes $A$, $B$ . The average gap is set to zero ($\Delta_{\vec{G}=0}=C=0$). The green dashed line shows the case of a gap landscape with \emph{realistic} amplitudes $A$, $B$ as given above. We see that the bands nearly coincide with the bands obtained for entirely ungapped graphene ($\Delta_i=0$, red solid line) and most importantly, no band gap opens. 
This holds even for unrealistically large values of the modulation amplitudes on the order of the hopping $t$ (also Fig. \ref{fig:bands}a, blue and purple dashed lines).
For large modulation amplitudes, another important feature becomes visible: a renormalization of the Fermi velocity $v_{\rm F}=\frac{1}{\hbar} \frac{\partial E}{\partial k}$. For $A$, $B$ being 100 times larger than the realistic values, $v_{\rm F}$ drops down by about 50\% (purple dashed line). For a realistic gap landscape, though, this effect is too small to be detected in experiments --- in contrast, e.g., to the case of twisted bilayer graphene \cite{twisted}. 

So, near the Fermi level, the energy bands of graphene are only weakly affected by a modulated gap landscape with realistic parameters and zero average gap -- no band gap opens and the amplitudes are too low to renormalize the Fermi velocity measurably. At the Brillouin zone boundary (Fig. \ref{fig:bands}b), however, a difference to the perfect isolated graphene becomes apparent: here, minigaps open similar to the case of graphene on Ir(111) moir\'es \cite{Ir_minigaps, Ir_ARPES}.

Now we discuss the scenario of a non-zero average gap, i.e., the case where the zeroth Fourier component is non-vanishing ($\Delta_{\vec{G}=0}=C \neq 0$). Fig. \ref{fig:bands2} shows the bands near the Fermi level of a \emph{realistic} gap landscape with $C=5$~meV (green dashed line) and $C=15$~meV (blue dashed line). Here, an absolute band gap on the order of $C$ opens that remains stable upon adding $\Delta_{\vec{G}\neq 0}$ terms of realistic amplitudes. Similar as for the Fermi velocity discussed in Fig. \ref{fig:bands} (a), our calculations showed that the band gap reduces measurably (but not entirely closes) when the modulation amplitude is increased by orders of magnitude. 
However, in the realistic scenario of graphene on h-BN, Fig. \ref{fig:bands2} clearly shows that the only quantity determining the absolute band gap is the zeroth Fourier component $\Delta_{\vec{G}=0}$.  
Hence, the average gap $\Delta_{\vec{G}=0}$ corresponds to the absolute spectral gap in the moir\'e structure, while amplitude and periodicity of spatially oscillating contributions to $\Delta$ renormalize the Fermi velocity. For the structure of Fig. \ref{fig:confs}b we find $\bar\Delta\approx-4$~meV. Therefore, we expect an absolute gap which is at least an order of magnitude smaller than the maximum local values of $|\Delta|$. This is well in line with the absence of a gap being reported in transport \cite{hBNmain} and STM experiments \cite{hBN_STM}. Our TB simulations further show that velocity renormalizations are below 2\% for the moir\'e structure of Fig. \ref{fig:confs}b.


\section{Real space shape of low energy states}
\label{sec:real_space}
\begin{figure*}
\includegraphics[width=0.93\columnwidth]{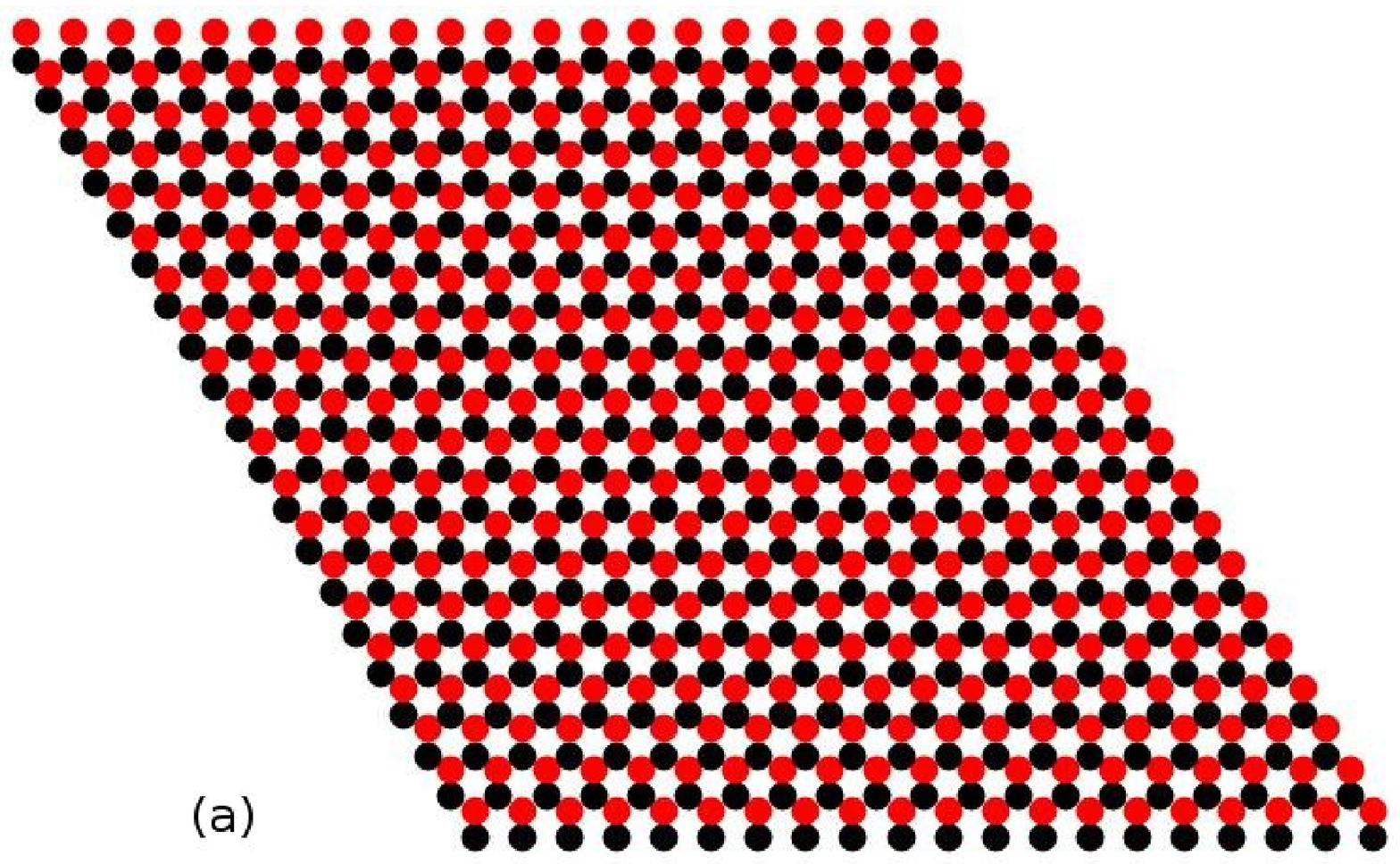}
\includegraphics[width=0.93\columnwidth]{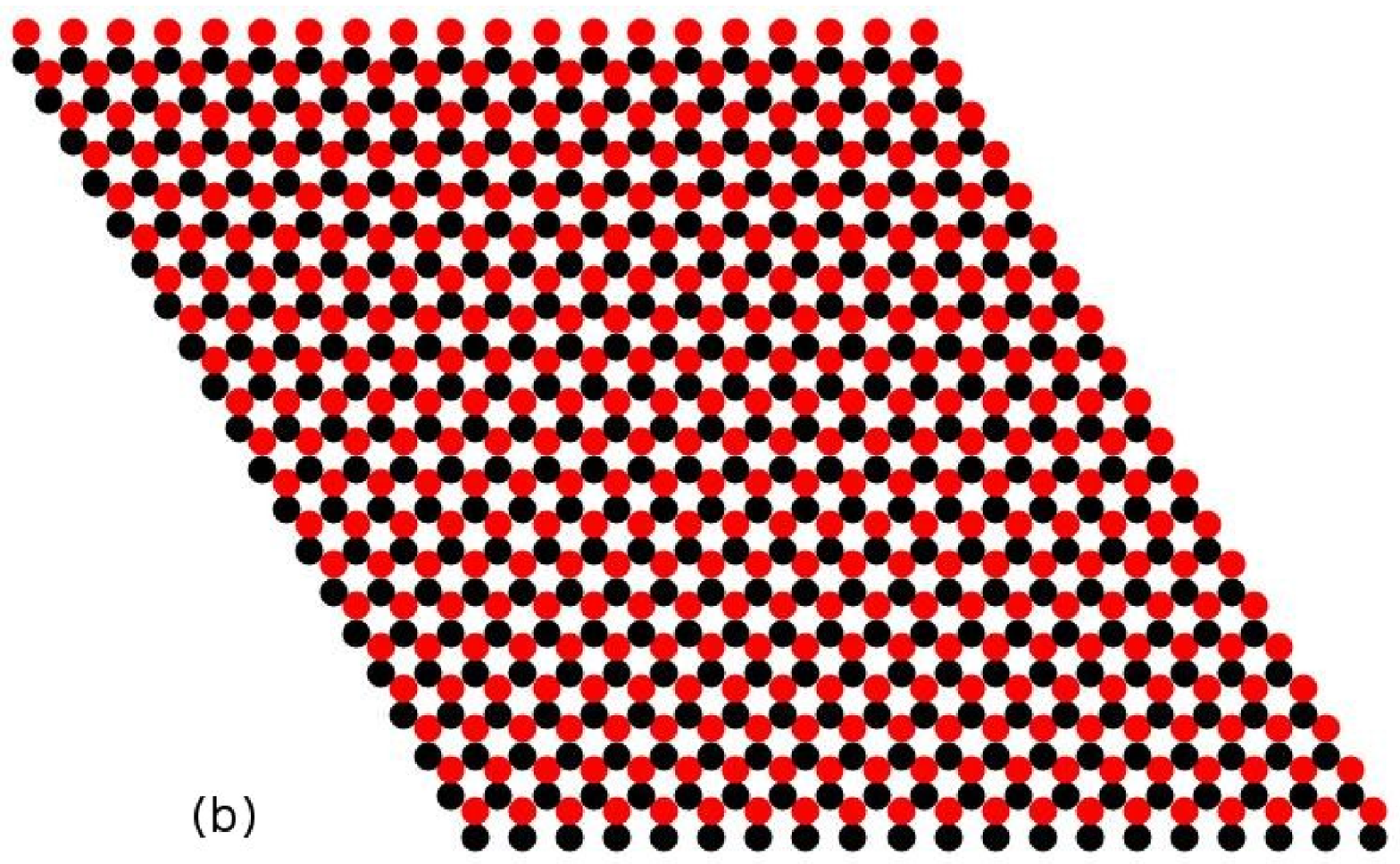}
\caption{\label{fig:realgaps} {(color online) 20x20 graphene supercell (red: sublattice A, black: sublattice B) with sinusoidally modulated gap terms with realistic amplitudes of $A=18.6$~meV, $B=42.0$~meV and vanishing average gap ($\Delta_{\vec{G}=0}=0$). The size of the dots depicts the contribution of each atom to states close to the Dirac point in an infinitesimal energy window around the valence band maximum (a) and the conduction band minimum (b).}} 
\end{figure*}

\begin{figure*}
\includegraphics[width=0.97\columnwidth]{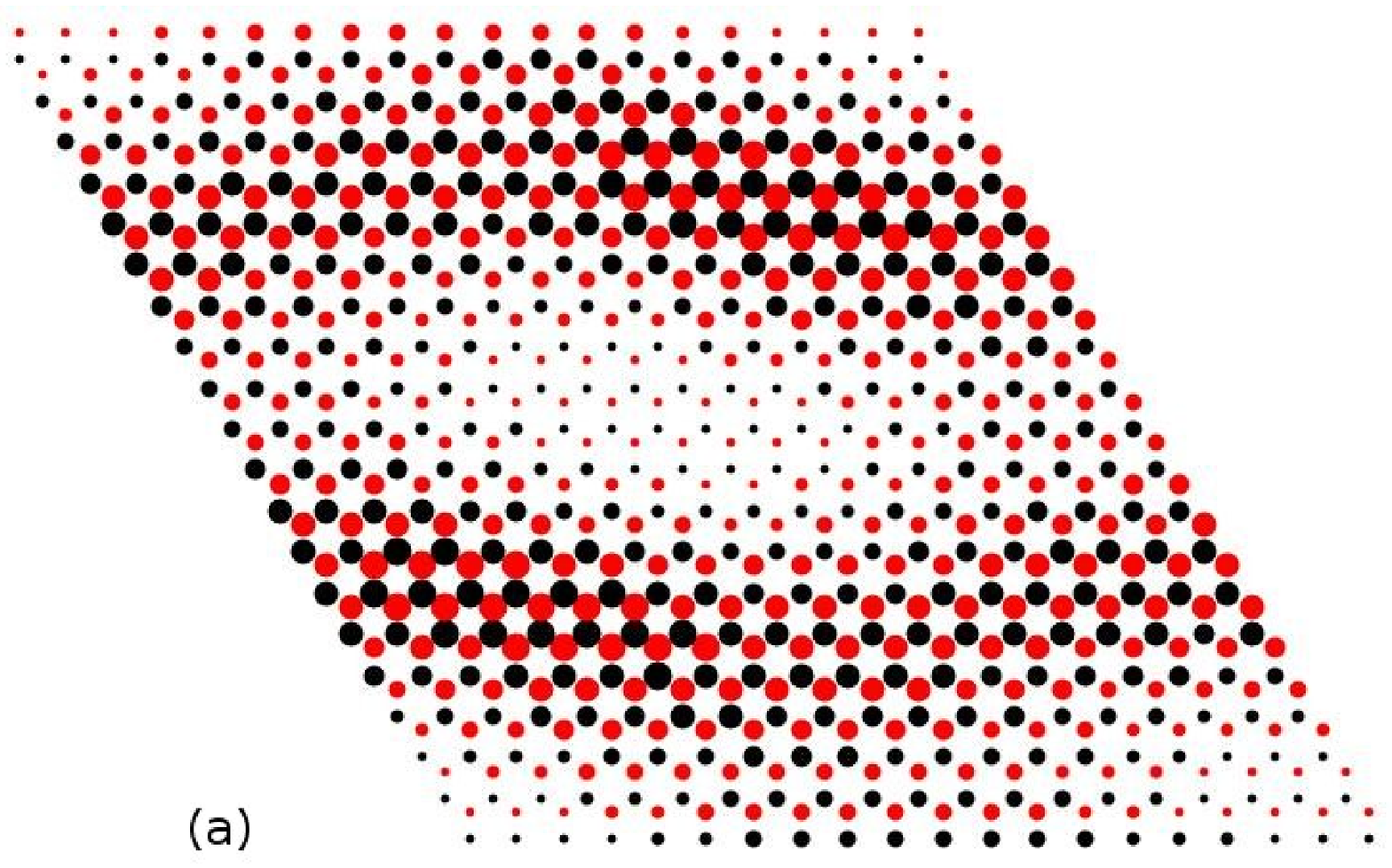}
\includegraphics[width=0.97\columnwidth]{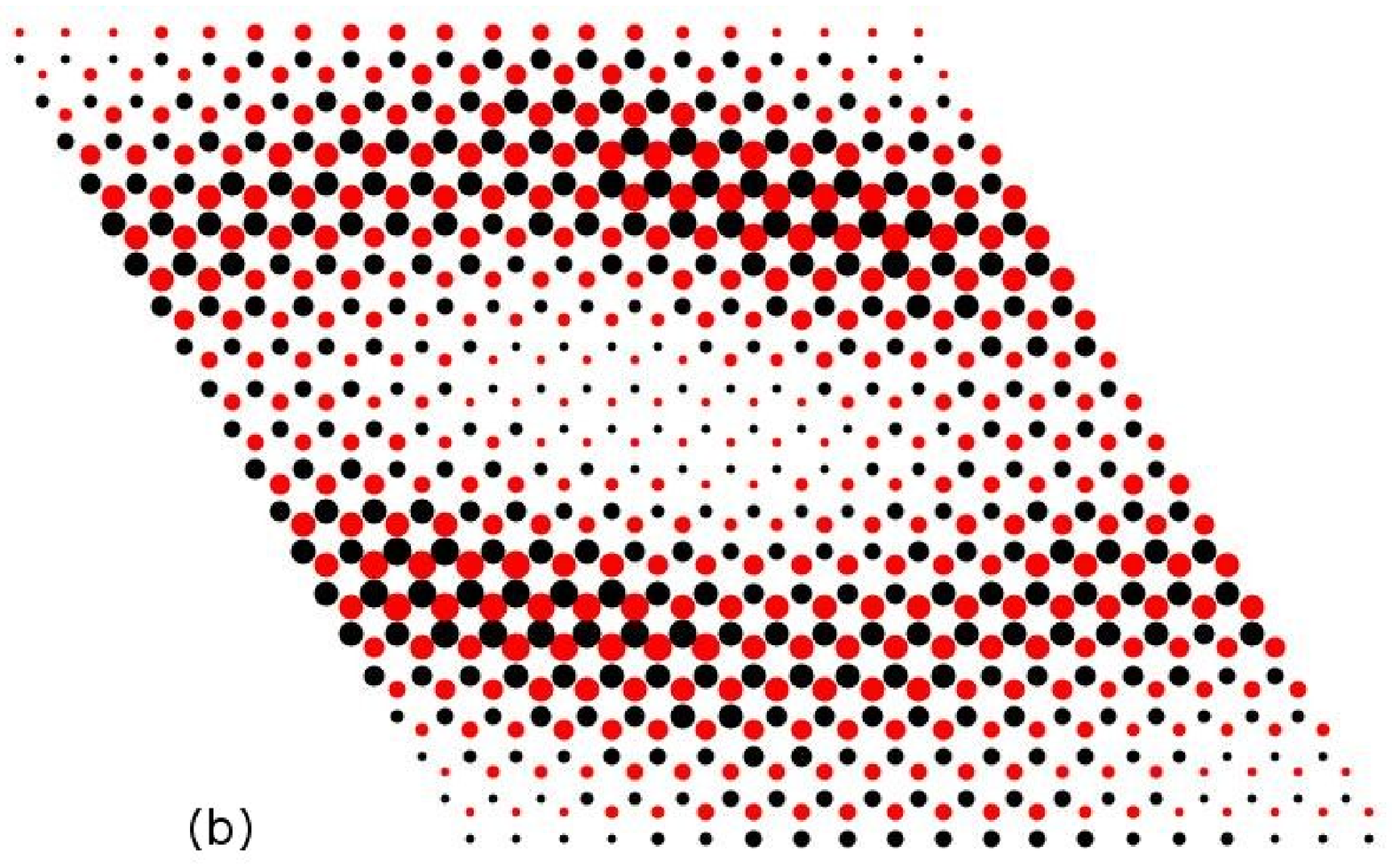}
\caption{\label{fig:sinusgap} {(color online) The same as Fig. \ref{fig:realgaps} for a 20x20 moir\'e cell with a sinusoidally modulated gap with amplitudes $\tilde A=25A$, $\tilde B=25B$, and zero average gap ($\Delta_{\vec{G}=0}=0$). So-called snake states occur.}}
\end{figure*}

\begin{figure*}
\includegraphics[width=0.97\columnwidth]{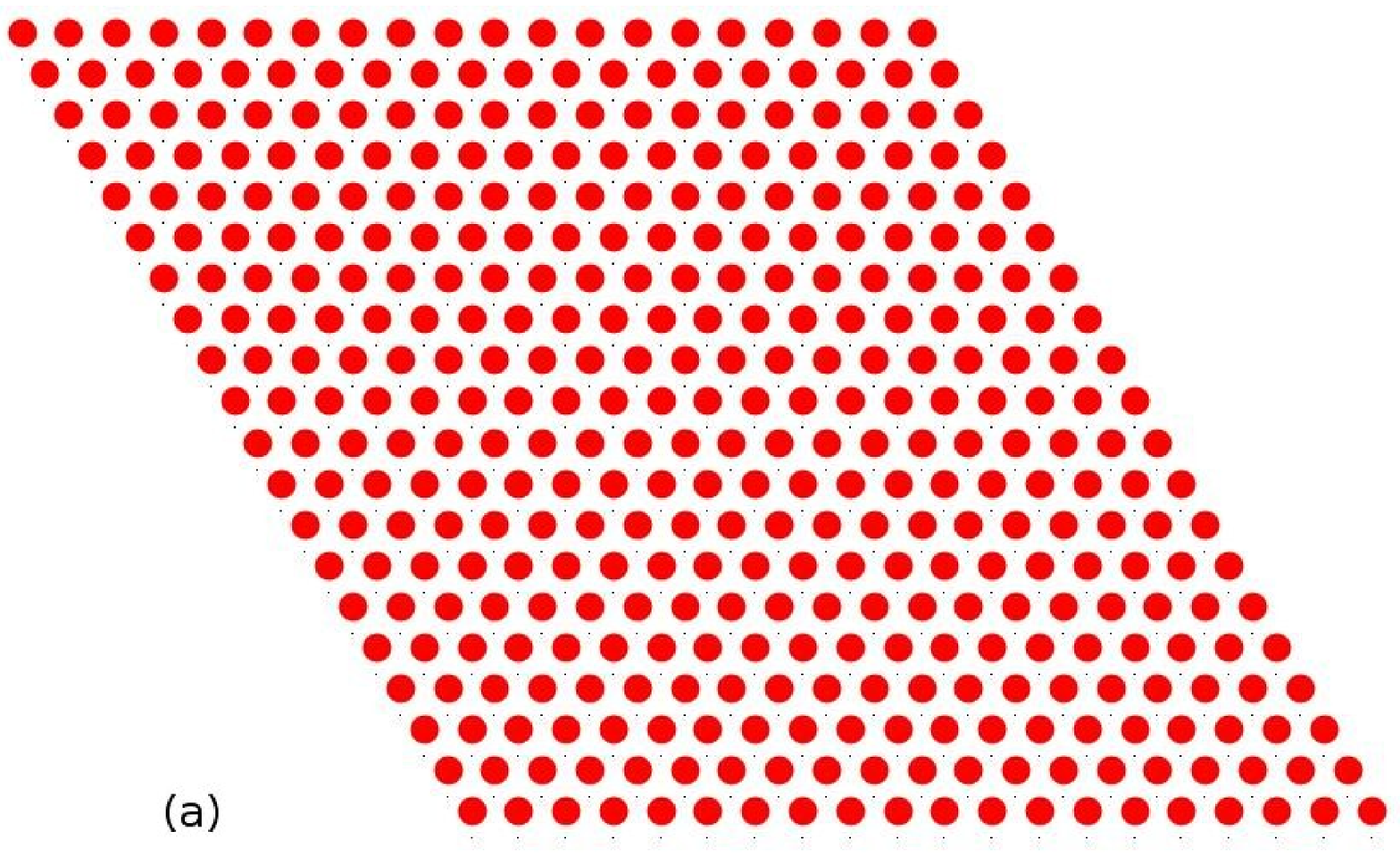}
\includegraphics[width=0.97\columnwidth]{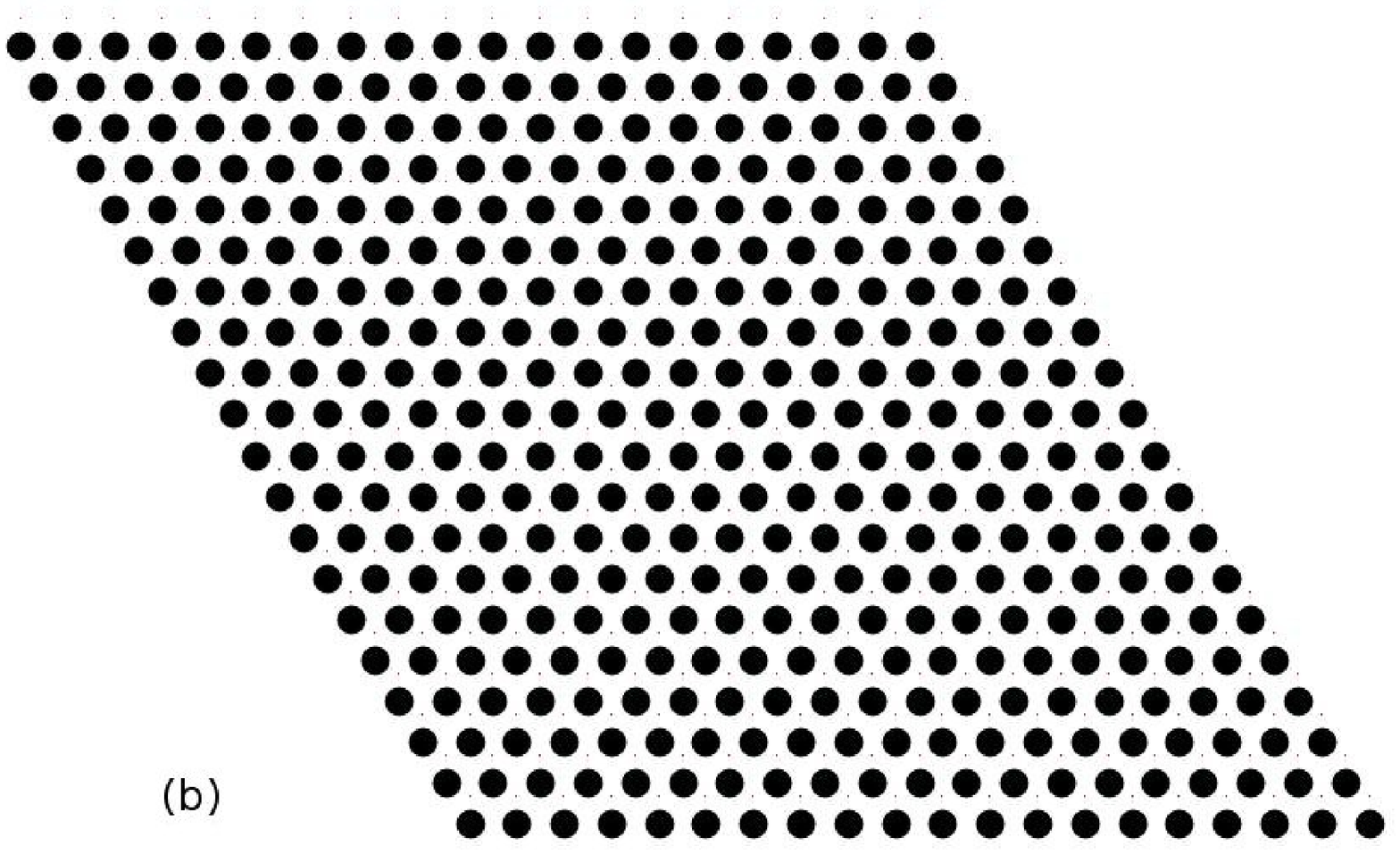}
\caption{\label{fig:homoggap} {(color online) The same as Fig. \ref{fig:realgaps} for a 20x20 cell of sinusoidally gapped graphene with realistic amplitudes $A$, $B$ and a finite average gap (here: $\Delta_{\vec{G}=0}=50$ meV).
}}
\end{figure*}

To understand how spatially modulated gap terms ($\Delta_{\vec{G}\neq 0}$) affect the graphene electrons and how they manifest, for instance, in local probe experiments, we visualize the states close to the Fermi level in real space. The Figs. \ref{fig:realgaps}, \ref{fig:sinusgap} and \ref{fig:homoggap} illustrate a 20x20 graphene supercell with sublattice A atoms as dots in black and sublattice B atoms as dots in red color. Here, the size of the dots illustrates the contribution of each atom to the states below the Fermi level (Figs. \ref{fig:realgaps}--\ref{fig:homoggap}a) and above (Figs. \ref{fig:realgaps}--\ref{fig:homoggap}b) in close proximity to the Dirac point. 

Fig. \ref{fig:realgaps} shows the case of realistic modulation amplitudes and vanishing average gap $\Delta_{\vec{G}=0}=0$. In this case, the amplitudes
of low energy states are equally distributed over both sublattices throughout the entire moir\'e cell as in ungapped graphene. There is no clear enhancement or decrease of probability density in any region of the moir\'e cell. 
However, increasing the gap modulation amplitude by a factor of 25 (Fig. \ref{fig:sinusgap}) induces a localization of the low energy states in regions where the local gap $\vert \Delta_i \vert$ is small --- so-called snake states occur, but no absolute band gap opens. Apparently, the states are equally localized in both sublattices. 

Whether or not snake states occur depends on the ratio of the modulation amplitudes $A,B$ to the energy $E_n\approx 2\pi\hbar v_F / (na)$ related to the moir\'e periodicity $na$. 20x20 and 50x50 moir\'e cells lead to $E_n\approx 0.7$~eV and $0.3$~eV, respectively. In the case of large modulation amplitudes $A,B > E_n$ (corresponing to Fig. \ref{fig:sinusgap}), the wave functions near the Dirac point clearly have the shape of snake states  --- in contrast to the case of $A,B \ll E_n$, where the low energy LDOS is almost homogeneous in the entire moir\'e cell (Fig. \ref{fig:realgaps}). The situation of Fig. \ref{fig:realgaps} is well in line with no gap or LDOS inhomogeneities being detected in the local probe experiments of Ref. \onlinecite{hBN_STM}. It is, however, important to note that a moir\'e periodicity of 100a which might be reached by external strain or twisting already corresponds to an intermediate case of $A$, $B$, and $E_n$ being on the same order magnitude.

The shape of the low energy states again changes for a system with a non-zero average gap $\Delta_{\vec{G}=0} \approx A,B$ 
on the order of the modulation amplitudes. In Fig. \ref{fig:homoggap} the case of a realistically modulated gap landscape and $\Delta_{\vec{G}=0}=+50$ meV is depicted: the states are homogeneously distributed in space fully but sublattice polarized -- with the state below Fermi level almost entirely localized in sublattice B (Fig. \ref{fig:homoggap} (a)) and vice versa (Fig. \ref{fig:homoggap} (b)). Our DFT calculations yield the case of a much smaller average gap on the order of few meV ($\Delta_{\vec{G}=0} \ll A,B$). In that case, two situations have to be distinguished: First, in the case of $\Delta_{\vec{G}=0} \ll A,B\ll E_n$ sublattice polarized states as in Fig. \ref{fig:homoggap} should be detectable in STM experiments, if the LDOS is measured inside an energies range $E<\Delta_{\vec{G}=0}$ which is sufficiently close around the Dirac point. Otherwise low energy states without any sublattice polarization as shown in Fig. \ref{fig:realgaps} will be detected. This situation would be again in line with the STM experiments of Ref. \onlinecite{hBN_STM}.

Differently, in the case of $\Delta_{\vec{G}=0} \ll E_n \ll A,B$, there is generally no the sublattice polarization detectable in the low energy LDOS but snake states similar to Fig. \ref{fig:sinusgap} occur.

\section{Conclusions}
\label{sec:concl}
In summary, we have calculated accurate adhesion energies for graphene--h-BN systems by means of ACFDT-RPA. 
A comparison of the strain energies with the adhesion energy differences suggests that a lattice mismatch between h-BN and graphene persists in experiments like Refs. \onlinecite{hBNmain, hBN_STM, ponomarenko2011tunable}. This explains the experimental observation of different stackings, moir\'e patterns, and stacking disorder. Our band structure calculations show that the gap landscape in the moir\'e structure exhibits mass terms with changing sign and a small average gap. 

The interplay of constant and spatially oscillating gap terms is decisive for determining whether or not phenomena like Anderson localization can occur \cite{randomgap} --- particularly in experiments where graphene is very close to the charge neutrality point \cite{ponomarenko2011tunable}. Gaps with spatially changing sign also control the transport properties of systems like (Hg,Cd)Te quantum well structures \cite{HgTe} which can be tuned into a topological insulator. While we find that structures like those in Refs. \onlinecite{hBNmain, hBN_STM, ponomarenko2011tunable} are likely not in a ``topological insulator regime'', where charge transport would occur through protected edge states, it remains to be seen whether this might be realized in structures like externally strained graphene on free-standing h-BN, where considerably larger moir\'e periodicities might be realized.


\begin{acknowledgments}
Support from the DFG (Germany) via SFB 668 and Priority Programme 1459 "Graphene", FOM (The Netherlands), and
computer time at HLRN (Germany) are acknowledged. We thank G. Kresse for helpful discussions.
\end{acknowledgments}

\appendix
\section{Elastic properties of \lowercase{h}-BN sheets}
\label{sec:elasticity}

Here, we now discuss the calculations of the elastic constants of h-BN (the elastic properties of graphene have been widely investigated in experiment \cite{grapheneyoung,chem_gra_strain_supp} and theory \cite{woelfle}). The isotropic Young's modulus and Lam\'e parameters of single h-BN sheets were obtained from DFT calculations. These elastic constants are determined by strong in-plane chemical bonds and well described within the LDA/GGA \cite{strainLDA_supp}. 
For single h-BN sheets, the two-dimensional Young's modulus is defined by
\begin{equation}
Y_{\rm 2D}= \frac{1}{A_0} \frac{\partial^2 E_{s}}{\partial \epsilon^2} \Big|_{\epsilon=0},
\end{equation}
where $\epsilon$ is the axial strain, $E_{s}$ the total strain energy and $A_0$ the equilibrium surface. The strain energies of a h-BN sheet were evaluated with uniaxial strains between -8\% (compression) and 8\% (tension). 
We obtain the two-dimensional Lam\'e parameters and Young's moduli of $\lambda=59$N/m, $\mu=125$N/m and $Y_{\rm h-BN,2D}=309$N/m within LDA and  $\lambda=54$N/m, $\mu=123$N/m, $Y_{\rm h-BN,2D}=300$N/m within GGA. The $Y_{\rm h-BN,2D}$ correspond to three-dimensional Young's moduli of  $Y_{\rm h-BN,3D}=0.94$TPa (LDA) and $Y_{\rm h-BN,3D}=0.91$TPa (GGA), assuming an interlayer separation of 3.3\AA~\cite{hBNlayersep}. Our full potential PAW calculations yield about 10\% higher elastic constants $Y_{\rm h-BN, 2D}$ and $\mu$ than calculations using Gaussian basis sets \cite{kudin_strain_supp}. The results are in agreement with the experiment, where $Y_{\rm h-BN,2D,exp}\approx 220 - 510$N/m was obtained for few-layer h-BN \cite{hBNyoung_supp}. 

\bibliography{hBNrefs}

\end{document}